\documentclass[12pt, preprint]{aastex}

\begin{document}

\title{Virialization in Dark Energy Cosmology}

\author{Peng Wang}

\affil{\it Department of Physics, Stanford University, Stanford, CA
94305-4060, USA} \email{pengwang@stanford.edu}

\begin{abstract}
We discuss the issue of energy nonconservation in the virialzation
process of spherical collapse model with homogeneous dark energy. We
propose an approximation scheme to find the virialization radius. By
comparing various schemes and estimating the parameter
characterizing the ratio of dark energy to dark matter at the
turn-around time, we conclude that the problem of energy
nonconservation may have sizable effects in fitting models to
observations.
\end{abstract}


\maketitle

\begin{center}
{\bf I. INTRODUCTION}
\end{center}

Analyzing the effects of dark energy on the nonlinear structure
formation process may provide us new ways of constraining the
properties of dark energy. Especially, a lot of recent works focused
on analyzing the effects of dark energy in the framework of the
spherical collapse model (\citealt{LLPR, Steinhardt, Maor, Mota,
Horellou, Battye, Iliev, Kamion, Nunes}). The spherical collapse
model is a simple but powerful framework to understand the growth of
bound systems in the universe \citep{Gunn}. It is also incorporated
in the famous Press-Schecter formalism \citep{PS}.

In spherical collapse model, we consider a top-hat spherical
overdensity with mass $M$ and radius $R$. At early times, it expands
along with the Hubble flow and density perturbations grow
proportionally to the scale factor. After the perturbation exceeds a
critical value, the spherical overdensity region will decouple from
the Hubble flow and go through three phases: (1) expansion to a
maximum radius, $R_{ta}$, after which the overdensity will
turn-around to collapse; (2) collapse; (3) virialization at the
virialization radius $R_{vir}$.

A key parameter of spherical collapse model is the ratio of the
virialized radius and turn-around radius $x=R_{vir}/R_{ta}$. Let's
first review briefly the derivation of the standard result $x=0.5$
in Einstein-de Sitter cosmology \citep{Peacock}.

As is well-known, the self-energy of a sphere of nonrelativistic
particles with mass $M$ and radius $R$ is
\begin{equation}
V_{mm}=-{3\over5}{GM^2\over R}\label{0}
\end{equation}

After the system virializes, the virial theorem $T_{vir}=(R/2)dU/dR$
and Eq.~(\ref{0}) tells us that $T_{vir}=-1/2U_{vir}$. Substituting
this to the energy conservation equation $T_{vir}+U_{vir}=U_{ta}$,
we get the standard result $x=0.5$.

When considering the evolution of spherical overdensities in the
presence of homogeneous dark energy (since \citet{Caldwell} have
shown that generally dark energy does not cluster on scales less
than 100 Mpc), the gravitational potential energy of the spherical
dark matter overdensity will be modified by a new term due to the
gravitational effects of dark energy on dark matter \citep{Maor,
Mota}:
\begin{equation}
U_{mQ}={1\over2}\int\rho_m\Phi_QdV \label{pe}
\end{equation}
where $\Phi_{Q}$ is the potential induced by dark energy
\begin{equation}
\Phi_Q=-{2\pi G}(1+3\omega_Q)\rho_Q \left(R^2-{r^2\over3}\right).
\label{poten}
\end{equation}

Note that if we do not consider the pressure of the dark energy, as
\citet{Steinhardt} did, then potential is proportional to $\rho$,
i.e. the factor $1+3\omega_Q$ should be unity in front of the
expression for $\Phi_{Q}$. In a fully relativistic treatment,
pressure will also contribute to gravitation, so the potential is
proportional to $\rho+3p$ \citep{Maor, Mota}.

Thus the total potential energy of spherical overdensity is
\begin{equation}
U=U_{mm}+U_{mQ}=-{3\over5}{GM^2\over R}-(1+3\omega_Q){4\pi
G\over5}M\rho_Q R^2,\label{totalpe}
\end{equation}
where we have substituted Eq.~(\ref{poten}) into Eq.~(\ref{pe}).

In most of the current literature \citep{Horellou, Battye, Iliev},
in the presence of smooth dark energy, $x$ is still found by using
the energy conservation equation
\begin{equation}
U_{vir}+T_{vir}=U_{ta}.\label{}
\end{equation}

Then from the virialization theorem
$T_{vir}={R_{vir}\over2}{dU(R)\over dR}\mid_{R=R_{vir}}$ and
Eq.~(\ref{totalpe}), we can find
\begin{equation}
T_{vir}=-{1\over2}U_{mm,vir}+U_{mQ,vir}.\label{T}
\end{equation}

Substituting Eq.~(\ref{T}) into the energy conservation equation,
one can find:
\begin{equation}
{1\over2}U_{mm,vir}+2U_{mQ,vir}=U_{mm,ta}+U_{mQ,ta},\label{}
\end{equation}
from which we can find the equation determining $x$
\begin{equation}
-4q(1+3\omega_Q)y^{-3(1+\omega_Q)}x^3+2[1+(1+3\omega_Q)q]x-1=0\label{old}
\end{equation}
where we have defined $q=\rho_{Q,ta}/\rho_{mc,ta}$ and $y=a_{vir}/
a_{ta}$. If we set the $(1+3\omega_Q)$ factor to unity in the first
two terms of Eq.~(\ref{old}), we can get the equation found by
\citet{LLPR, Steinhardt}.

\begin{center}
{\bf II. THE ENERGY NON-CONSERVATION PROBLEM}
\end{center}

The procedure described in Sec. I is problematic when dark energy is
dynamical, i.e. $\omega_Q\neq-1$. Indeed, when $\omega_Q>-1$,
$\rho_Q$ is decreasing with time. When considering collapse of dark
matter halo of the cluster scale, since dark energy does not cluster
below 100 Mpc \citep{Caldwell}, $\rho_Q$ in $U_{mQ}$ should take its
background value, i.e. evolving with time. In other words, dark
energy does not virialize with dark matter, otherwise it cannot be
smooth. Thus, $U_{mQ}$ will contribute a non-conservative force to
the dark matter particle. So the clustering dark matter with
potential (\ref{totalpe}) is a non-conservative system:
\begin{equation}
U_{vir}+T_{vir}< U_{ta}.\label{}
\end{equation}

So actually, in the presence of dark energy, dark matter cannot
reach virialization in the strict sense. But for dark matter halo of
the cluster scale, it clusters at the era when the effect of dark
energy is still small. So it is reasonable to assume that dark
matter particles can reach a quasi-equilibrium state in which virial
theorem holds instantaneously. This is supported by the observations
of relaxed cluster in our Universe (see e.g. \citet{allen}). In the
following discussion, we still call this quasi-equilibrium state as
virialization. Thus assuming dark matter has reached this
quasi-equilibrium state, its total energy can be computed by the
virial theorem,
\begin{equation}
U=U_{vir}+{R\over2}{dU\over R}\mid_{R=R_{vir}}=-{3\over10}{GM^2\over
R_{vir}}-(1+3\omega_Q){4\pi
G\over5}MR_{vir}^2\rho_{Q,ta}\left(a(t)\over
a_{ta}\right)^{-3(1+\omega_Q)},\label{qq}
\end{equation}
which is decreasing with time. Although this non-conservation effect
is small in our discussion, it is worth commenting that in dark
energy dominated era, this effect may be large. For example, in the
extreme case of phantom dark energy models, the effect of dark
energy may be so large that cluster, galaxy and even our solar
system will de-virialize in the future \citep{kamion2}.

So using Eq.~(\ref{old}) to find $x$ will generally overestimate its
actual value. In fact, from the dark matter potential
(\ref{totalpe}), we can see that $U_{total,
vir}=U_{mm,vir}/2+2U_{mQ,vir}$ is a monotonically increasing
function of $x$. Thus, while in fact we have $U_{total,
vir}<U_{total, ta}$, if we still use $U_{total,vir}=U_{total,ta}$ to
determine $x$, we will get a $x$ larger than its actual value.

If the dark energy density is time-independent, then the system with
the potential (\ref{totalpe}) is conservative and thus we can use
energy conservation legitimately. Thus to estimate the virialization
radius when dark energy density is changing with time, we can take
$\rho_{Q,vir}$ to be the same as $\rho_{Q,ta}$.

With this approximation, the equation determining $x$ is,
\begin{equation}
-4q(1+3\omega_Q)x^3+2[1+(1+3\omega_Q)q]x-1=0.\label{main}
\end{equation}
It is worth commenting that taking $\rho_{Q,vir}$ to be
$\rho_{Q,ta}$ does not mean ignoring the background evolution of
dark energy, i.e. make it degenerate with a true cosmological
constant. First, there is a factor $1+3\omega_Q$ in Eq.~(\ref{main})
which is different from the case of a cosmological constant. Second,
the value of $\rho_{Q,ta}$ is different from $\rho_{Q,0}$; while for
a true cosmological constant, $\rho_{\Lambda}$ is constant all the
times. Thus using Eq.~(\ref{main}) to estimate $x$ can give us a
more realistic value than Eq.~(\ref{old}) and at the same is able to
discriminate among different values of $\omega$.

\begin{figure}
  \epsscale{.80}
  \plotone{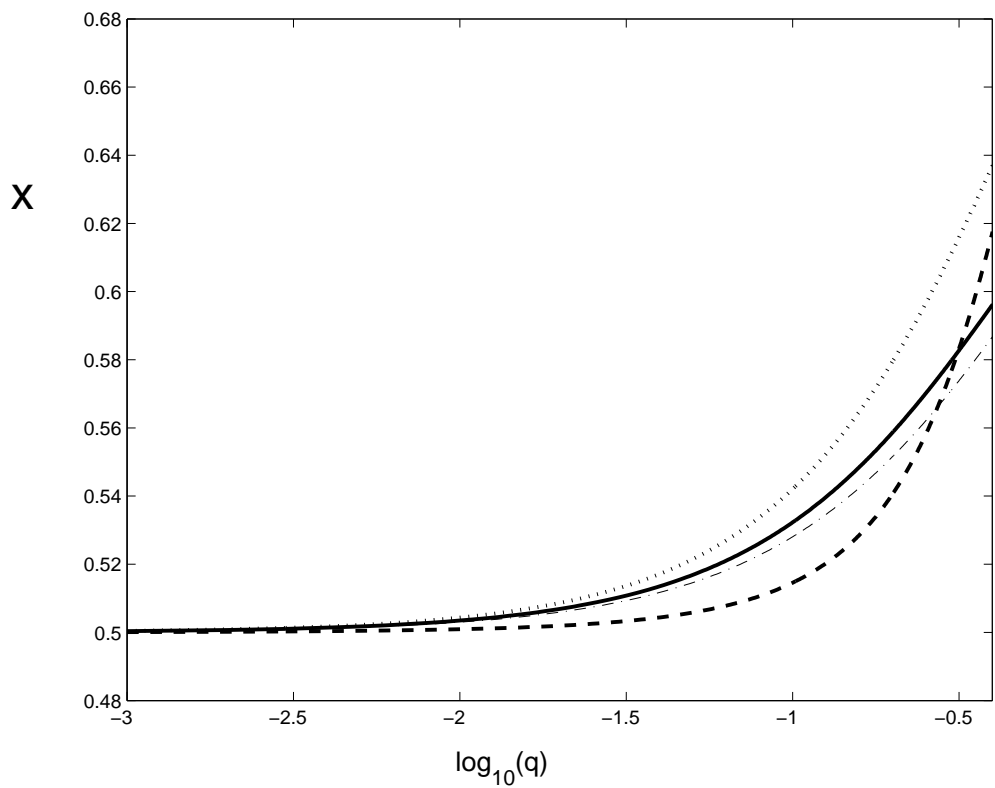}
    \caption{Ratio of the virialization radius to the turn-around radius $x=R_{vir}/R_{ta}$
    as a function of $q$, which characterizes the strength of dark energy at turn-around, for
    $\omega_Q=-0.8$. The dotted line is computed by Eq.~(\ref{old}), the solid one is computed by Eq.
    (\ref{main}), the dashed line is computed by Eq.~(28) of ~\citet{Maor} and the dashed-dotted line is computed by Eq.~(\ref{1.1})}\label{1}
\end{figure}

Figure 1 shows the virialization radius to the turn-around radius
$x=R_{vir}/R_{ta}$ as a function of $q$.  From the figure, it can be
seen that when $q$ is large, i.e. the effects of dark energy is
large, Eq.~(\ref{old}) (dotted line) will always predict a larger
virialization radius than Eq.~(\ref{main}) (solid line). This showed
explicitly the comment following Eq.~(\ref{qq}). Thus by assuming
the dark energy density to be constant during the virialization
process, we can use energy conservation and we can find a lower $x$
which is closer to the actual one.

It is also interesting to note that all the curve in Fig.~1 is above
the standard value $x=0.5$ in an Einstein-de Sitter universe. This
is easy to understand. Since dark energy will cause an effective
repulsive force on the dark matter, the dark matter particles can
reach equilibrium with a larger radius. Thus, $x>0.5$ is a smoking
gun of dark energy (see also \citet{Maor}, which reaches similar
conclusion).

Recently, \citet{Maor} considered the possibility that even if dark
energy does not fully cluster, it still fully virialize. Then there
will also be a energy non-conservation problem because the
virialized system (now containing both dark matter and dark energy)
does not cluster in the same rate. Note that this energy
non-conservation problem is different from the energy
non-conservation problem discussed above. In \citet{Maor}'s
analysis, since both dark matter and dark energy virialize, they
included the dark energy self-energy when using the virial theorem.
In our case, where dark energy is smooth and do not virialize, the
physical picture of what's going on is just a spherical overdensity
of dark matter particles collapsing in the background of smooth dark
energy. From the derivation of the virial theorem \citep{Marion}, we
know that the potential energy appearing in the virial theorem is
the one that will give rise to forces on the virialized particles.
So if dark energy does not virialize, we should consider only the
potential energies giving rise to dark matter self-gravitation and
its gravitational interaction with dark energy. This is why our
energy corrected equation (\ref{main}) is very different from that
of \citet{Maor} (Eq.~(28) in it).

In Fig. 1, we showed the prediction of the virialization equation
found by \citet{Maor} (dashed line). It can seen that for $q$ not
too large, the prediction of \citet{Maor} is much smaller than our
result Eq.~(\ref{main}). This is conceivable. Since in
\citet{Maor}'s analysis, the positive self-energy of dark energy is
also included in the total potential energy of virialized particles,
$U_{total}$ will be larger than its actual value. Since $U_{total}$
is a monotonically increasing function of $x$, with the same initial
energy, the equation of \citet{Maor} will thus predict a smaller
$x$.

\citet{Mota} have considered the case of fully clustered and
virialized dark energy. We think it is interesting to consider the
case that only a portion of dark energy cluster and virialize. In
this case, we should include the dark energy self-energy that will
cluster since this part of the self-energy will contribute to the
force felt by the virialized dark energy particles.

In this case, the dark energy evolution equation is
\begin{equation}
\dot\rho_Q+(1-F)3{\dot R\over R}(1+\omega_Q)\rho_Q+F3{\dot a\over
a}(1+\omega_Q)\rho_Q=0,\label{}
\end{equation}
where $F$ characterizes the fraction of dark energy that cluster.
This equation can be integrated to find
\begin{equation}
\rho_Q=\rho_{Q,ta}x^{-3(1+\omega)(1-F)}y^{-3(1+\omega)F}\label{}
\end{equation}

First, we use our approximation method: we neglect the background
evolution of the dark energy, i.e., we take
\begin{equation}
\rho_Q=\rho_{Q,ta}x^{-3(1+\omega)(1-F)}y^{-3(1+\omega)F}\rightarrow
\rho_{Q,ta}x^{-3(1+\omega)(1-F)},\label{}
\end{equation}
in the virialization process. Taking into account the observation
that we should only include a fraction $1-F$ of the dark energy
potential energy, we can get the equation determining $x$
\begin{eqnarray}
&&-(1-F)(1+3\omega_Q)\left[7-6(1+\omega_Q)(1-F)\right]q^2x^{-6\omega_Q+6F(1+\omega_Q)}\cr
&&-(2+3\omega_Q-F)[4-3(1+\omega_Q)(1-F)]qx^{-3\omega_Q+3F(1+\omega_Q)}\cr
&&+2[1+(1+3\omega_Q)q+(1-F)q+(1-F)(1+3\omega_Q)q^2]x-1=0\label{gamma}
\end{eqnarray}

For $F=1$, Eq.~(\ref{gamma}) will reduce to Eq.~(\ref{main}), while
for $F=0$, it will reduce to the equation found by \citet{Mota}.
Thus, our result Eq.~(\ref{main}) can be continuously connected to
the case that dark energy will also collapse with dark matter. This
is physically satisfying.

Second, if we adopt the proposal of restoring energy conservation by
\citet{Maor} then when $F=1$, the virialization equation is
\begin{equation}
-2(1+3\omega_Q)qx^{-3\omega_Q}-2(1+3\omega_Q)qy^{-3(1+\omega_Q)}x^3+[1+(1+3\omega_Q)q]x-1=0\label{1.1}
\end{equation}

We showed the dependence of $x$ on $q$ from Eq.~(\ref{1.1}) as the
dashed-dotted line in Fig. 1. It can be seen that although the
approach of restoring energy conservation are different in Eqs.
(\ref{1.1}) and (\ref{main}), their predictions are rather close.
This illustrates that although the underlying ideas are different,
in practice, our approximation scheme is quantitatively close to the
scheme of \citet{Maor}. In both cases, the difference from the old
result (\ref{old}) is large when $q$ is large.

From Fig. 1 we can also see that for $q\sim10^{-2}$ or smaller, we
get $x=0.5$ in all the four approaches. This is reasonable. In the
virialization process, it is the self-energy of matter that plays
the dominant role. In fact, $U_{QQ.vir}/U_{mQ,vir}$ and
$T_{QQ,vir}/T_{mQ,vir}$ are both of the order
\begin{equation}
(1+3\omega_Q)q\left({a_{vir}\over
a_{ta}}\right)^{-3(1+\omega_Q)}\left({R_{vir}\over
R_{ta}}\right)^3\label{}
\end{equation}
since $\left({a_{vir}\over a_{ta}}\right)^{-3(1+\omega_Q)}\simeq
1.6^{-3(1+\omega_Q)}$ and $\left({R_{vir}\over R_{ta}}\right)^3\sim
0.1$, for $q<0.01$ the above ratio is much smaller than $1$, and
thus we can expect that for small $q$, the problem of energy
conservation will not influence the virialization process greatly.

\begin{figure}
  \epsscale{.80}
  \plotone{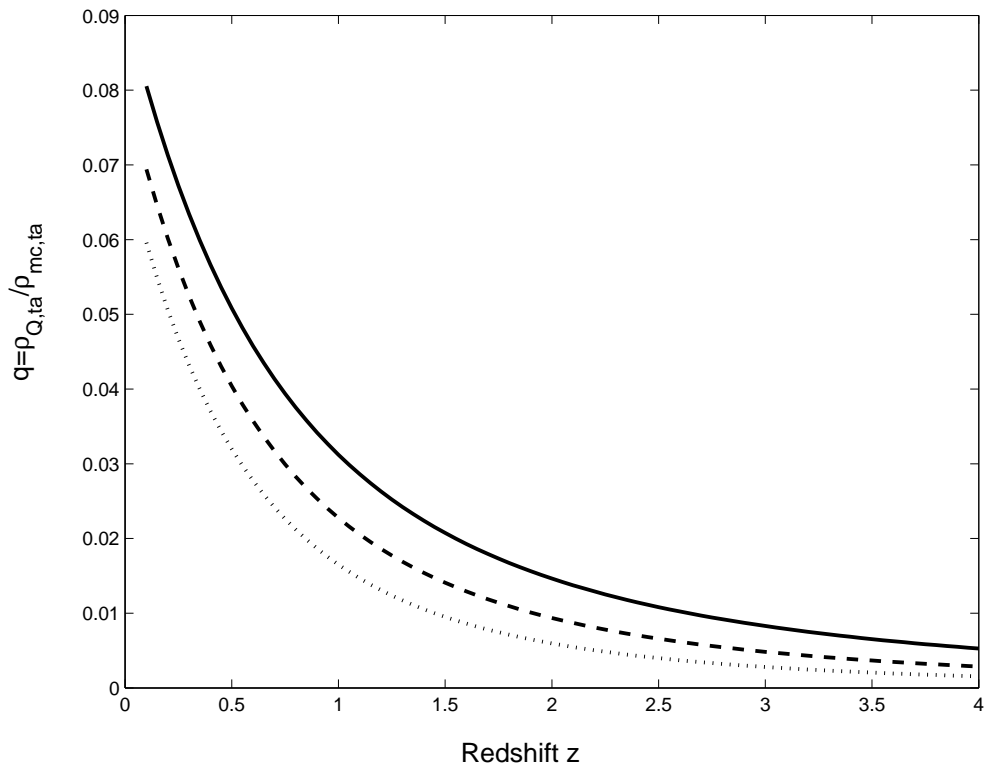}
    \caption{The dependence of $q$ on $z_{vir}$ from Eq.~(\ref{q}). The solid, dashed, dotted lines correspond to
$\omega_Q=-0.7,-0.8,-0.9$, respectively.}\label{1}
\end{figure}

Thus to estimate the effects of dark energy on virialization, and
especially the ambiguity of energy-nonconservation, it is necessary
to estimate the value of $q$ for the virialization redshift
$z_{vir}$ that would be interesting to observations. If for
observationally interesting $z_{vir}$, $q$ will always be quite
small, then we can conclude that the problem of energy
non-conservation will not bother us too much in analyzing the
effects of dark energy on the formation of non-linear structure.
Unfortunately, this is not the case.

Let's begin by writing $q$ as
\begin{equation}
q={\rho_{Q,ta}\over\rho_{mc,ta}}={\rho_{Q,ta}\over
\zeta\rho_{m,ta}}={\Omega_{Q0}(1+z_{ta})^{3\omega_Q}\over\zeta\Omega_{m0}},\label{q}
\end{equation}
where we have defined $\zeta=\rho_{mc,ta}/\rho_{m,ta}$.

First, after specified $z_{vir}$, $z_{ta}$ can be computed using the
fact that $t_{vir}=2t_{ta}$,  which is due to the observation that
collapse proceeds symmetrically to the expansion phase. Then, we use
the fitting formula for $\zeta$ presented by \citet{Steinhardt}:
\begin{equation}
\zeta=\left({3\pi\over4}\right)^2\Omega_{m,ta}^{-0.79+0.26\Omega_{m,ta}-0.06\omega_Q}.\label{}
\end{equation}

With those two inputs, we can get the dependence of $q$ on $z_{vir}$
from Eq.~(\ref{q}) shown in Fig. 2. It can be seen that $q$ will be
of the order $10^{-2}$ when the virialization redshift is larger
than roughly $2$. Combining this with Fig. 1, we conclude that for
observationally interesting clusters, i.e. clusters formed after
redshift $2$, the presence of dark energy will have sizable
modifications to the standard result $x=0.5$. Thus observational
evidence for $x>0.5$ would be a strong evidence in favor of
dynamical dark energy.

\begin{figure}
  \epsscale{.80}
\plotone{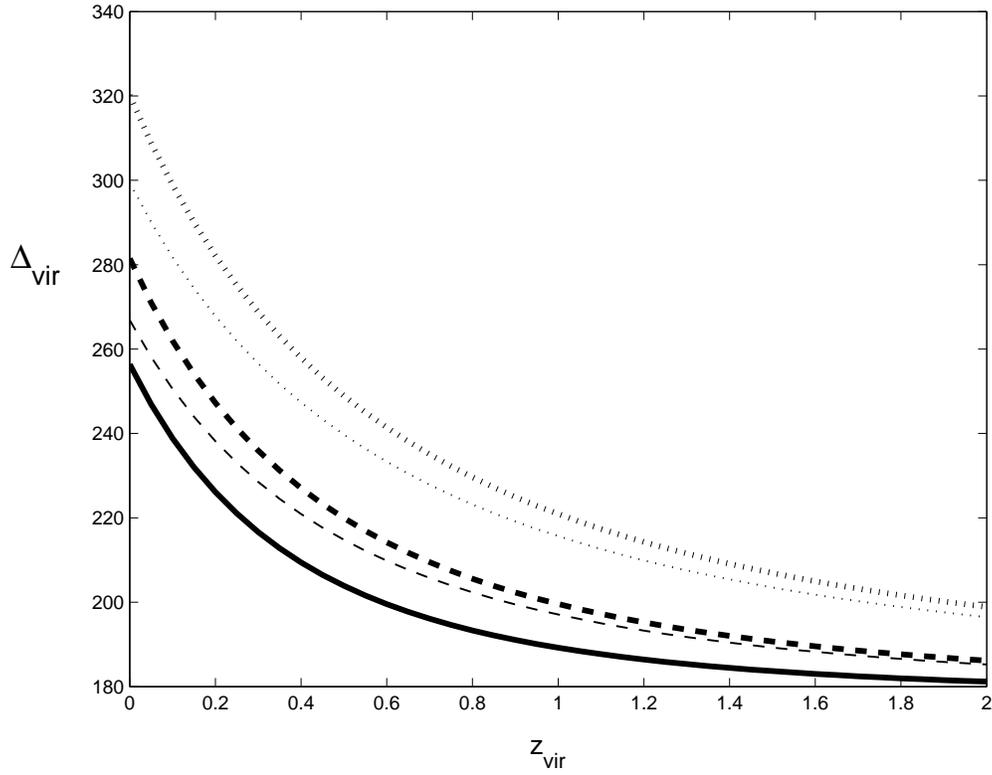}
    \caption{Thick lines: the dependence of $\Delta_{vir}$ on $z_{vir}$ for
$\omega_Q=-0.6,-0.8, -1$ from top to bottom using Eq.~(\ref{main});
thin lines: the dependence of $\Delta_{vir}$ on $z_{vir}$ for
$\omega_Q=-0.6,-0.8$ from top to bottom using Eq.~(\ref{old}), the
case of $\omega=-1$ is identical with that of using
Eq.~(\ref{main}). }\label{1}
\end{figure}

An important parameter in fitting theoretical calculations to
observation is the density contrast at virialization
$\Delta_{vir}\equiv\rho_{mc,vir}/\rho_{m,vir}=\zeta y^3x^{-3}$. Fig.
3 shows the dependence of $\Delta_{vir}$ on $z_{vir}$ for
$\omega_Q=-0.6,-0.8, -1$ from top to bottom using Eq.~(\ref{main})
and Eq.~(\ref{old}). We can see that for $\omega_Q$ not too close to
$-1$, there is obvious differences in the predicted $\Delta_{vir}$
using Eq.~(\ref{main}) and Eq.~(\ref{old}). Thus, whether including
the effect of energy conservation may have large impact on fitting
models to cosmological observation such as weak lensing
\citep{Kamion}. It is worth commenting that if we ignore the
$1+3\omega$ factor in $U_{mQ}$, as in \citet{Steinhardt, Kamion},
the difference will be even larger. For example, \citet{Kamion}
computed that for $\omega_Q=-0.6$, $\Delta_{vir}\sim 420$ for
$z_{vir}=0$. Furthermore, as can be seen in Fig. 3, whatever scheme
we use to find $x$, there is notable difference between the case of
a true cosmological constant and dynamical dark energy. Thus,
cluster observations may provide important information on the
dynamical behavior of dark energy.

\begin{center}
{\bf III. CONCLUSIONS AND DISCUSSIONS}
\end{center}

To summarize, in this work we discussed the issue of energy
non-conservation in the virialzation process of spherical
overdensity with homogeneous dark energy. We proposed that taking
the dark energy density to be constant during the virialization
process to obtain an estimate of the virialization radius. By
comparing various schemes and estimating the parameter $q$, we
conclude that there will be sizable effect of dark energy on
virialization process. A general signature of dark energy is that
the final virialization radius will be larger than half of the
turn-around radius.

It should be emphasized that the analysis in this work is quite
qualitative. More detailed numerical simulations and analysis of
observational data are required to estimate quantitatively the
effect of dark energy on spherical collapse models and answer the
general question ``can we constrain the evolution of dark energy by
studying the structures of non-linear objects in our Universe".
Furthermore, establishing firmly the result $x>0.5$ from observation
is challenging. In practice, it is much easier to measure directly
baryons in clusters. But there are some astrophysical processes
leading to energy non-conservation in the virialization process of
baryon in the dark matter halo (e.g. X-ray emission of the hot gas,
conduction, AGN heating, dynamical friction, etc., see e.g.
\citet{Kamion2} and references therein). We should compare the
effects of those processes to dark energy in realistic analysis. To
achieve this, we need to know the concrete physical mechanism of
virialization in both the dark matter and baryon sectors, which is
now still not well-understood. Thus more works in this direction is
needed and will be rewarding.

\begin{center}
{\bf ACKNOWLEDGEMENT}
\end{center}
I would like to thank Irit Maor for helpful comment on the first
version of this paper.

\end{document}